\documentstyle[preprint,aps]{revtex}
\input epsf.tex
\def\DESepsf(#1 width #2){\epsfxsize=#2 \epsfbox{#1}}
\begin{document}
\pagestyle{empty}                                      
\preprint{
\font\fortssbx=cmssbx10 scaled \magstep2
\hbox to \hsize{
\hbox{
            hep-ph/9809282}
\hfill $
\vtop{
 \hbox{ }}$
}
}
\draft
\vfill
\title{
Prospects for Direct $CP$ Violaton in
Exclusive and Inclusive Charmless B decays}
\vfill
\author{$^{1,2}$Xiao-Gang He, $^{1,3}$Wei-Shu Hou, and $^4$Kwei-Chou Yang}
\address{
\rm $^1$Department of Physics, National Taiwan University,
Taipei, Taiwan 10764, R.O.C.\\
\rm $^2$ School of Physics, University of Melbourne,
Parkville, Vic. 3052, Australia\\
\rm $^3$ Physics Department, Brookhaven National Laboratory,
Upton, NY 11973, USA\\
and\\
\rm $^4$Institute of Physics, Academia Sinica, Taipei, R.O.C}

%
%
\vfill
\maketitle
\begin{abstract}
Within the Standard Model,
$CP$ rate asymmetries for $B\to K^-\pi^{+,0}$ could reach 10\%.
With strong final state phases, they could go up to 20--30\%,
even for $\bar K^0\pi^-$ mode which would have opposite sign.
We can account for $K^-\pi^{+}$, $\bar K^0\pi^-$ and $\phi K$
rate data with new physics enhanced color dipole coupling and
destructive interference.
Asymmetries could reach 40--60\% for $K\pi$ and $\phi K$ modes
and are all of the same sign.
We are unable to account for $K^-\pi^0$ rate.
Our inclusive study supports our exclusive results.
\end{abstract}
%
%
\pacs{PACS numbers: 13.20.He, 11.30.Er  12.60.-i
 }
%
%
\pagestyle{plain}


Half a dozen charmless two body $B$ decays appeared in 1997 \cite{TwoBody},
suggesting that loop induced $b\to s$ penguin processes are prevalent.
Recently, the $\bar K^0 \pi^-$ rate has come down, and the $K^-\pi^0$ mode
has just been observed \cite{Kpi0}.
Together with $K^-\pi^+$, all three modes
are now $\simeq 1.4\times 10^{-5}$, with error bars of order 30\%.
The limit on the pure penguin mode $B\to \phi K < 0.5\times 10^{-5}$,
however, is rather stringent.
At present, one has ${\cal O}(10)$ events per observed mode.
As B Factories turn on at SLAC and KEK
and with the CLEO III upgrade at Cornell,
these numbers should increase
to ${\cal O}(10^2)$ per experiment in two years, and
to ${\cal O}(10^3)$ within five years at the B Factories.
Many new modes would also emerge.
Equally crucial, one would finally have event by event $K/\pi$ separation.
Thus, {\it direct} $CP$ violating rate asymmetries ($a_{CP}$)
at $30\%$ and down to $10\%$ levels can be
probed in the above time frame.
It is of importance to know whether such large $a_{CP}$s
are possible, and, if observed, whether they
would signal the presence of new physics.

Within the Standard Model (SM),
$a_{CP}$ for $b\to s$ modes are suppressed by  \cite{GH}
${\rm Im}\,(V_{us}V_{ub}^*)/(V_{cs}V_{cb}^*) \simeq\eta\lambda^2 < 1.7\%$,
where $\lambda \cong \vert V_{us} \vert$ and $\eta < 0.36$ is
the single $CP$ violating parameter in the Wolfenstein parametrization
of the Kobayashi-Maskawa (KM) matrix.
Asymmetries need not be small, however,
when both tree and penguins contribute,
such as for $\bar B^0 \to K^-\pi^+$ mode,
where destructive interference could lead to
$a_{CP} \sim 10\%$.
If final state interaction (FSI) phases are present,
$a_{CP}$s could even reach beyond 20\%.
But, as we will show,
a combined study of several modes can distinguish
between FSI phases or the presence of new physics.
In contrast, pure penguin $b\to s\bar ss$ modes
have only one single amplitude, and within SM
the $\sim \eta\lambda^2$ suppression cannot be evaded.
They are hence sensitive probes of new physics phases.
The observation of $a_{CP}$ above the 10\% level in
these modes would be striking evidence
for physics beyond SM.

We shall study both exclusive and inclusive $a_{CP}$s for
charmless $b(p_b) \rightarrow s(p_s) \bar q(p_{\bar q}) q(p_q)$ decays,
starting from the effective Hamiltonian
\begin{eqnarray}
H_{\rm eff} &=& {4G_F\over \sqrt{2}} \left[
V_{ub}V_{us}^*(c_1O_1 + c_2 O_2)
-V_{ib}V_{is}^* \, c^i_j O_j\right],
\label{Heff}
\end{eqnarray}
where $i$ is summed over $u,c,t$ and $j$ is summed over $3$ to $8$,
with operators defined as
\begin{eqnarray}
O_1 &=& \bar u_\alpha \gamma_\mu L b_\beta \,
        \bar s_\beta \gamma^\mu L u_\alpha,
\;
O_2 = \bar u \gamma_\mu L b \, \bar s \gamma^\mu L u,
\;
O_{3,5} = \bar s \gamma_\mu L b \, \bar q \gamma^\mu {L(R)} q,
\nonumber\\
O_{4,6} &=& \bar s_\alpha\gamma_\mu L b_\beta \,
      \bar q_\beta \gamma^\mu {L(R)} q_\alpha,
\;\;\;
\tilde O_8 = {\alpha_s\over 4\pi}\, \bar s i\sigma_{\mu\nu} T^a
                {m_b q^\nu\over q^2} Rb\, \bar q \gamma^\mu T^a q,
\nonumber
\end{eqnarray}
where $\tilde O_8$ arises from the dimension 5 color dipole $O_8$ operator,
and $q= p_b-p_s$.
We have neglected electroweak penguins for simplicity.
The Wilson coefficients $c_j^i$ are evaluated to NLO order
in regularization scheme independent way,
for $m_t = 174$ GeV, $\alpha_s (m_Z^2) = 0.118$ and $\mu = m_b = 5$ GeV.
Numerically  \cite{desh-he},
$c_{1,2} = -0.313,\ 1.150$, $c_{3,4,5,6}^t = 0.017,\ -0.037,\ 0.010,\ -0.045$,
and $c_8^{\rm SM} = c_8^t-c_8^c= -0.299$.
For absorptive parts,
we add
$c^{u,c}_{4,6}(q^2) = -Nc_{3,5}^{u,c}(q^2)=-P^{u,c}(q^2)$
for $c$ and $u$ quarks in the loop, where
$8\pi P^{u,c}(q^2) = \alpha_s c_2 (10/9 + G(m_{u,c}^2,q^2))$,
and
%
$G(m^2,q^2) = 
4\int x(1-x)\, dx\, {\rm ln}\, (m^2/\mu^2 - x(1-x)q^2/\mu^2).
$
To respect CPT and unitarity at ${\cal O}(\alpha_S^2)$,
one must properly \cite{GH} include
the absorptive part of the gluon propagator for the $b\to s\bar u u$ mode.
Hence, for $c^t_{3-8}$ at $\mu$ below $m_b$, we substitute
$4\pi\,{\rm Im}\, c_8(q^2) = \alpha_s c_8
                      \sum_{u,d,s,c} {\rm Im}\, G(m_i^2,q^2)$,
and
%
$8\pi\, {\rm Im}\, c^t_{4,6} = -8\pi N\, {{\rm Im}\, c^t_{3,5}}
={\alpha_s} [{c^t_3}\, {\rm Im}\, G(m_s^2,q^2)
+ (c^t_4+c^t_6)
    \sum_{u,d,s,c}{\rm Im}\, G(m_i^2,q^2)]$,
but only when these interfere with
the tree amplitude.

We use the $\tilde O_8$ operator to illustrate the
possibility of new physics induced $a_{CP}$.
Although $b\to sg$
(with $g$ ``on-shell" or jet-like) is only $\sim$ 0.2\% in SM,
data actually still allows \cite{bsglimit}
it to be $\sim 5\%$--$10\%$,
which would help alleviate \cite{bsg} the long-standing
low charm counting and semileptonic branching ratio problems.
The recent discovery of \cite{etapXs} a surprisingly large
semi-inclusive charmless $B\to \eta^\prime + X_s$ decay
could also be \cite{HT} hinting at $\vert c_8 \vert \sim2$.
Such a large dipole coupling would naturally carry a
{\it KM-independent} $CP$ violating phase,
$c_8 = \vert c_8\vert e^{i\sigma}$, and $a_{CP} \sim$ 10\%
in the $m_{X_s}$ spectrum of $B\to \eta^\prime + X_s$
is possible.
It is clear that {\it such a new phase
could lead to large $a_{CP}$ in a plethora of $b\to s\bar qq$ modes}
 \cite{Kagan}.

The theory of exclusive rates is far from clean.
One needs to evaluate all possible hadronic matrix elements
of products of currents.
Faced with recent CLEO data,
many theorists have advocated \cite{Neff} the
use of $N_{\rm eff}\neq  3$ as a process dependent
measure of deviation from factorization,
which becomes a mode by mode fit parameter.
One still has to assume form factors and pole values,
and, for $a_{CP}$ evaluation, the $q^2$ value to take.
The latter is further clouded by FSI phases.
Even with such laxity, there are problems \cite{Neff}
already in accounting for observed rates.
The $\eta^\prime K$ and $\omega K$ modes
appear to be high, while
the yet to be observed $\phi K$ mode seems too low and
hard to reconcile with large $\eta^\prime K$.
We refrain from studying $B\rightarrow \eta^\prime K$
as it probably has much to do with the anomaly mechanism.

Our central theme is whether large $a_{CP}$ is possible
in $b\to s$ modes,
and, if so, how would they signal the presence of new physics,
such as enhanced $c_8$.
We find that SM alone allows for sizable $a_{CP}$ in $K\pi$ type of modes.
This is important for the early observability of $a_{CP}$s, 
so let us first investigate the $K\pi$ modes.

The $K^-\pi^+$ mode receives
both tree and penguin $b\to s\bar uu$ contributions,
hence we separate into two isospin amplitudes, $A = A_{1/2} +  A_{3/2}$.
Since color allowed amplitudes dominate,
we take $N_{\rm eff} \simeq N = 3$.
Assuming factorization we find,
\begin{eqnarray}
A_{1/2} &=&
i{G_F\over \sqrt{2}} f_K F_0\; (m_B^2-m_\pi^2)
\left \{V_{us}^*V_{ub} \left[ {2\over 3} \left({c_1\over N} + c_2\right)
-{r\over 3}\left(c_1+{c_2\over N}\right) \right] \right .
\nonumber\\
& - &V_{js}^*V_{jb}  \left. \left[ {c_3^j\over N} + c_4^j
+{2m_K^2\over (m_b-m_u)(m_s+m_u)}\left({c_5^j\over N} +c_6^j\right)
+ \delta_{jt} {\alpha_s\over 4\pi} {m_b^2\over q^2} c_8 \tilde S_{\pi K}\right] \right\},
\label{kpi1}
\end{eqnarray}
and for $A_{3/2}$ \cite{A3half} one sets $c^j_{3-8} \to 0$
and $2/3,\ -r/3 \to 1/3,\ r/3$.
Here, $F_0 = F_0^{B\pi}(m^2_K)$ is a BSW form factor,
$\tilde S_{\pi K} \sim -0.76$ is a complicated form factor normalized to $F_0$
coming from the matrix element of $\tilde O_8$
(with further assumptions),
and $r = f_\pi F^{BK}_0(m^2_\pi) (m_B^2 - m_K^2)
/f_K  F_0^{B\pi}(m_K^2) (m_B^2 - m_\pi^2)$.
The $K^- \pi^0$ mode is similar,
with changes in  $A_{3/2}$ and an overall factor of $1/\sqrt{2}$.
Since penguins contribute only to $A_{1/2}$,
for $B^- \to \bar K^0 \pi^-$ one has 
just Eq. (\ref{kpi1}) with $c_{1,2} \to 0$.
Note that the $c_{5,6}$ terms are sensitive to
current quark masses
because of effective density-density interaction.
The impact of the $c_8$ term is small in SM.

The absorptive parts for $c^j_{3-8}$ is evaluated at
some $q^2$ for the virtual gluon.
We take $q^2 \approx m_b^2/2$
which favors large $a_{CP}$, but
it could be as low as $m_b^2/4$ \cite{GH}.
In usual convention,
the dispersive part of second term of Eq. (\ref{kpi1}) is negative,
while the sign of first term depends on
$\cos\gamma$ (or, Wolfenstein's $\rho$)
where $\gamma = {\rm arg}\, (V^*_{ub}V_{us})$.
For $\cos\gamma > 0$ which is now favored,
$a_{CP}$ is enhanced by destructive interference,
but for $\cos\gamma < 0$ the effect is opposite \cite{GH}.
This can be seen in Fig. 1(a) and (b) where we
plot the branching ratio ($B_r$) and $a_{CP}$ vs. $\gamma$.
For $K^- \pi^{+,0}$, $a_{CP}$ peaks at the sizable $\sim 10\%$
just at the currently favored \cite{stocchi} value of
$\gamma \simeq 64^\circ$.
But for $\bar K^0 \pi^-$, $a_{CP} \sim \eta\lambda^2$ is very small.
We have used $m_s(\mu = m_b) \simeq 120$ MeV \cite{koide}
since it enhances the rates.
With $m_s(\mu = 1$ GeV) $\simeq$ 200 MeV,
the rates would be a factor of 2 smaller.
We find $K^- \pi^+$, $\bar K^0 \pi^-$, $K^- \pi^0
 \sim 1.4,\ 1.6,\ 0.7 \times 10^{-5}$, respectively.
The first two numbers agree well with experiment \cite{Kpi0},
but $K^- \pi^0/K^- \pi^+ \sim (1/\sqrt{2})^2$ seems too small.

As noted some time ago \cite{GH}, the $K\pi$ modes are sensitive
to FSI phases of the two isospin amplitudes.
If the FSI phase difference $\delta$ is large, it could easily overhwelm
the meager {\it perturbative} absorptive parts controlled by ``$q^2$".
Neglecting an overall phase,
we write $A = A_{1/2} +  A_{3/2} e^{i\delta}$
and plot in Fig. 1(c) and (d) the $B_r$ and $a_{CP}$ vs. $\delta$ for
$\gamma = 64^\circ$.
The rate is not sensitive to $\delta$ which reflects
penguin dominance over tree, but
$a_{CP}$ can now reach $20\%$, even reaching 30\% for $K^-\pi^0$.
We stress that the $\bar K^0 \pi^-$ mode
is in fact also quite susceptible to FSI phases,
as it is the isospin partner of $K^-\pi^0$
which does receive tree contributions.
When $\delta \neq 0$, tree contributions enter
$B^- \to \bar K^0 \pi^-$ through FSI rescattering.
Comparing Fig. 1(b) and (d), $a_{CP}$ in this mode can be
{\it much larger} than naive expectations.
However, being out of phase with $K^-\pi^+$, 
comparing the two cases can give information on $\delta$.

To illustrate physics beyond SM, we keep $N = 3$
but set $c_8 = 2 e^{i\sigma}$.
Since the $c_8$ term now dominates,
the results are not very sensitive to the FSI phase $\delta$.
We plot in Fig. 1(e) and (f) the $B_r$ and $a_{CP}$
vs. the new physics phase $\sigma$,
for $\gamma = 64^\circ$ and $\delta = 0$.
The $K^-\pi^+$ and $\bar K^0 \pi^-$ modes are very close in rate
for $\sigma \sim 45^\circ - 180^\circ$,
but the $K^-\pi^0$ mode is still a factor of 2 too low.
However,
the $a_{CP}$ can now reach 50\% for $K^-\pi^+$/$\bar K^0 \pi^-$
and 40\% for $K^-\pi^0$!
These are truly large asymmetries and would be easily observed soon.
They are in strong contrast to the SM case with FSI phase $\delta$,
Fig. 1(d), and can be distinguished.

Genuine pure penguin processes arising from $b\to s\bar ss$
give cleaner probes of new physics $CP$ violation effects.
The amplitude for $B^-\to \phi K^-$ decay is
\begin{eqnarray}
A(B\rightarrow \phi K)&\simeq&
-i {G_F} f_\phi m_\phi
                       \sqrt{2} p_B\cdot \varepsilon_\phi F_1(m_\phi^2)\, V_{js}^*V_{jb}
\left\{ \left(c_{3}^j+{c_4^j/N} + c_5^j\right) \right|_{q_2^2}
\nonumber\\
 && + \left.\left({c_3^j/N} +c_4^j + {c_6^j/N}\right)\right|_{q_1^2}
        + \left.
 \delta_{jt} {\alpha_s} {m_b^2} c_8 \tilde S_{\phi K}/ 4\pi q_1^2 \right\}.
\label{phiK}
\end{eqnarray}
The annihilation contributions which we have neglected 
are small, and the tree annihilation contributions
are further suppressed by the KM factor. 
We have also dropped the color octet contributions 
and have checked that they are small.
The relevant $q^2$ is determined by kinematics:
$q_1^2 = m_b^2/2$ as before,
but for amplitudes without Fierzing $q_2^2 = m_\phi^2$.
Since the amplitude is now basically pure penguin, 
$c_8$ should have no absorptive part.
As seen from Fig. 1(a) and (b),
the SM rate of $\sim 1\times 10^{-5}$ is a bit high 
while $a_{CP}$ is unmeasurably small.
For $c_8 = 2 e^{i\sigma}$,
the results are plotted in Fig. 1(e) and (f) vs. $\sigma$.
The rate is lower than the $\bar K^0 \pi^-$/$K^-\pi^+$ modes
because it is not sensitive to $1/m_s$.
The $a_{CP}$ could be as large as 60\%
when $c_8$ and the SM amplitude interfere destructively.

One can now construct an attractive picture.
As commented earlier, recent studies cannot  explain
the low  $B\to \phi K$ upper limit.
If $c_8$ is enhanced and interferes destructively with SM,
$B\to \phi K$ can be brought down to below $5\times 10^{-6}$.
The experimentally observed $\bar K^0 \pi^- \simeq K^-\pi^+$
follows from $c_8$ dominance,
and the $B_r$ value $\simeq 1.4\times 10^{-5}$ which is 2--3 times larger
than the $\phi K$ mode suggests a low $m_s$ value and
slight tunings of BSW form factors.
Around $\sigma \sim 145^\circ$,
the rates are largely accounted for,
but $a_{CP}$ for $\phi K$, $K^-\pi^+/K^-\pi^0$ and $\bar K^0 \pi^-$
could be enhanced to the dramatic values of
55\%, 45\% and 35\% respectively.
We do fail to account for the $K^-\pi^0$ rate,
which is comparable to the $\phi K$ mode.
We note that,
with recalibration of CLEO II data and 
adding similar amount of CLEO II.5 data,
the $\bar K^0 \pi^-$ rate came down by 40\% \cite{Kpi0}!
Thus,
a $K^-\pi^0$ rate lower than the present preliminary result
cannot be excluded.
If the current result of
$\bar K^0 \pi^- \simeq K^-\pi^+ \simeq K^-\pi^0 \simeq 1.4\times 10^{-5}$
persists down to errors of say $15\%$,
the enhanced $c_8$ model would be in trouble.
From Fig. 1(a) and (c) we see that SM with Eq. (\ref{Heff}) also
does not suffice, even with FSI phases,
and other effective interactions such as electroweak penguins have to be
included.

We are barely able to accommodate $B\to \omega K$.
Within SM $1/N_{\rm eff} \sim 1$ is needed to
account for $B\to \omega K \simeq 1.5\times 10^{-5}$,
while for $1/N_{\rm eff} \sim 0$ one accounts for at most only half.
With $c_8 = 2 e^{i\sigma}$,
we can account for $B_r$ for both large and small $N_{\rm eff}$,
but not for $N = 3$. However, $a_{CP}$ is never more than a few \%
and not very interesting.

To gain better understanding,
we discuss briefly inclusive $b\to s\bar qq$ decays,
where the theory is cleaner.
The existence of $b\to sg$ at lower order implies
a $\log q^2$ pole for the $\vert c_8\vert^2$ term,
which we simply cut off at $q^2 \simeq 1$ GeV$^2$.
The pure penguin $b\to s\bar dd$ case is the simplest
since $c_{1,2}$ do not contribute.
The results for SM and several $|c_8| = 2 $ cases are
given in Fig. 2(a) and Table 1.
The low $q^2$ pole is not prominent and $a_{CP}$ is indeed small in SM,
arising mostly from below $\bar cc$ threshold.
For enhanced $c_8$, however, both the low $q^2$ pole and
the $a_{CP}$ above $\bar cc$ threshold become significant.
The overall $a_{CP}$ is not much larger than SM case,
since the $\bar uu$ cut and hence the asymmetry
below $\bar cc$ threshold is still suppressed by $V_{us}^*V_{ub}$.
However, above the $\bar cc$ threshold $a_{CP}$
is of order 10--30\%, which
confirms our exclusive findings, and
perhaps can be probed by the partial reconstruction technique
developed in Ref. \cite{Browder}.
We note that for the destructive interference case of $\sigma = 3\pi/4$,
the rate is comparable to SM and $a_{CP}$ is the largest.

The $b\to s\bar uu$ process receives tree contributions also.
Keeping $c_1$ and $c_2$ in the calculation,
the results are given in Fig. 2(b) and Table 1.
We now have to include gluon propagator absorptive parts
for $c_{3-8}^t$ terms when they interfere with tree amplitude.
As noted in  Ref. \cite{GH}, in SM
the $a_{CP}$ tends to cancel between $q^2$ below
and above $\bar cc$ threshold,
but in each domain the $a_{CP}$ could be of order 10\%,
supporting our $B \to K^- \pi^{+,0}$ studies.
For $|c_8| = 2$, rather large $a_{CP}$ can arise
above the $\bar cc$ threshold.
For $b\to s\bar ss$ (Fig. 2(c)) mode,
interference with exchange graphs from identical particle effects
leads to peculiar shapes
and smears out the rate asymmetry to all $q^2$, but
the qualitative features are similar to the $b\to s\bar dd$ case.
Our inclusive results therefore provide qualitative support of our
exclusive studies.

We conclude that the prospects are rather bright for observing
large $CP$ violating asymmetries in charmless $b\to s$ decays in the near future.
Within SM, $a_{CP}\sim 10\%$ for $K^-\pi^+$ and $K^-\pi^0$
for $\gamma \sim 64^\circ$ which is currently favored,
but $< 1\%$ for $\bar K^0\pi^-$ and $\phi K$.
With large FSI phase $\delta$, $a_{CP}$ in $K\pi$ modes
can be enhanced to 20--30\%,
even for the naively pure penguin $\bar K^0\pi^-$,
but the latter would typically have sign opposite
to $K^-\pi^+$/$K^-\pi^0$.
Enhanced color dipole $c_8 \sim 2 e^{i\sigma}$
could explain $\bar K^0\pi^- \sim K^-\pi^+$,
which are split upwards from the $\phi K$ mode by
a low $m_s$ value.
Destructive interference with $\sigma \sim 145^\circ$
seems to be favored by present data on $K^-\pi^+$, $\bar K^0\pi^-$ and
$\phi K$.
The corresponding $a_{CP}$ would be $\sim 35$--$45\%$ for $K\pi$ modes
and $55\%$ for $\phi K$, which should be easily observed
and rather distinct from SM case with or without FSI phase.
We are unable to account for the newly observed $K^-\pi^0$ rate.
Noting that the $\bar K^0\pi^-$ rate came down recently,
we wait for further confirmation of present data.
Our inclusive study of rates and asymmetries support our exclusive findings.
Our results on large $a_{CP}$
in charmless $b\to s$ decays can be tested soon at the B Factories.

This work is supported in part by
grants NSC 88-2112-M-002-033 and NSC 88-2112-M-001-006
of the Republic of China, and by Australian Research Council.

\begin{table}
\caption{Inclusive $B_r$ (in $10^{-3}$)/$a_{CP}$ (in \%) for SM and for $c_8 = 2 e^{i\sigma}$.}
\begin{tabular}{cccccccc}
&SM &$\sigma$ =\ \ \ \ \ \  0 \ \ \ \ \ & $\frac{i\pi}{4}$ & $\frac{i\pi}{2}$ & $\frac{i3\pi}{4}$
 &$i\pi$  \\
\hline $b\to s \bar d d$ & 2.6/0.8   & \ \ \ \ \ \ 8.5/0.4 & 7.6/3.4 &
5.2/6.5 & 2.9/8.1 & 1.9/0.5 \\ \hline
\hline $b\to s \bar u u$ & 2.4/1.4   & \ \ \ \ \ \ \ 8.1/-0.2 & 7.5/2.6 &
5.5/5.6 & 3.2/8.1 & 2.0/3.5 \\ \hline
\hline $b\to s \bar s s$ & 2.0/0.9 & \ \ \ \ \ \ 6.9/0.4 & 6.2/3.2 &
4.4/6.0 & 2.6/7.1 & 1.8/0.4 \\ \hline
\end{tabular}
\end{table}

\begin{figure}[htb]
\centerline{ 
\DESepsf(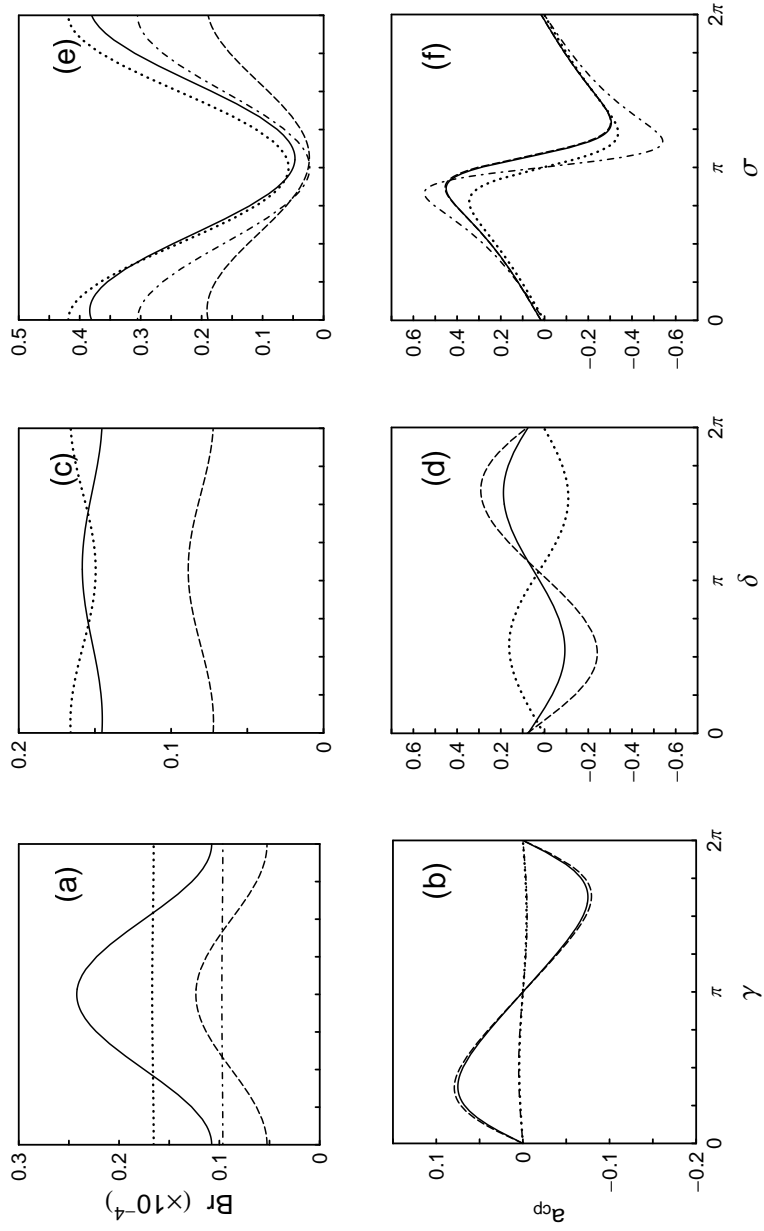 width 10 cm) } \vskip2cm
\caption { 
$Br$ and $a_{CP}$ vs. (a), (b) SM unitarity angle $\gamma$, 
(c), (d) FSI phase $\delta$ for $\gamma = 64^\circ$, and 
(e), (f) new physics phase $\sigma$ for $\gamma = 64^\circ$ and $\delta = 0$. 
Solid, dotted, dashed and dotdashed lines are for 
$K^-\pi^+$, $\bar K^0 \pi^-$, $K^-\pi^0$ and $\phi K$ respectively.
} 
\label{fig1}
\end{figure}

\begin{figure}[htb]
\centerline{ 
\DESepsf(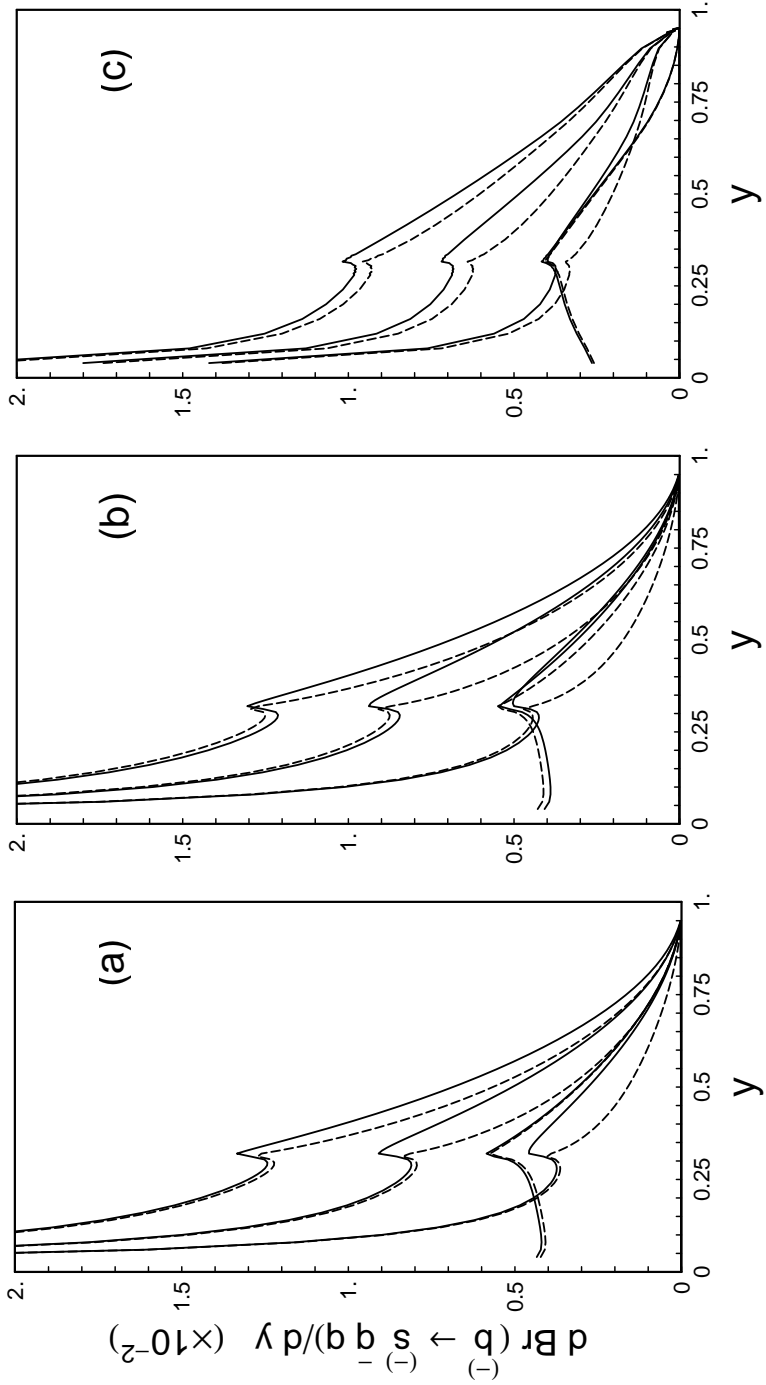 width 10 cm) } \vskip1cm 
\caption {
Inclusive branching ratios vs. $y = q^2/m_b^2$ for
(a) $b\to s\bar dd$, (b) $b\to s\bar uu$ and (c) $b\to s\bar ss$ decays
(solid) and $\bar b$ decays (dashed). The curves with
prominent small $y$ tail are for 
$c_8 = 2e^{i\sigma}$ with 
$\sigma = \pi/4$ (top), $\pi/2$ (middle), $3\pi/4$ (bottom), 
while the other is the SM result. 
} 
\label{fig2}
\end{figure}

\end{document}